\documentclass[conference]{IEEEtran}

\usepackage{cite}
\usepackage{array}
\usepackage{tabularx}
\usepackage{amsmath,amssymb,amsfonts}
\usepackage{graphicx}
\usepackage{tikz}
\usetikzlibrary{calc,arrows.meta,positioning}
\usepackage{subcaption}
\usepackage{booktabs}
\usepackage{multirow}
\usepackage{url}
\usepackage[hidelinks]{hyperref}

\makeatletter
\g@addto@macro\UrlBreaks{\do-\do\_}
\makeatother
\newcommand{\E}{\mathbb{E}}
\newcommand{\Var}{\mathrm{Var}}

\title{Diagnosing High-Performance BFT Consensus via Mixture Modeling of Block Time Distributions}

\author{
\IEEEauthorblockN{Hongru He}
\IEEEauthorblockA{
Chiba Institute of Technology\\
2-17-1 Tsudanuma, Narashino, Chiba 275--0016, Japan\\
hehongru5@gmail.com
}
\and
\IEEEauthorblockN{Akihiro Fujihara}
\IEEEauthorblockA{
Chiba Institute of Technology\\
2-17-1 Tsudanuma, Narashino, Chiba 275--0016, Japan\\
akihiro.fujihara@chibatech.ac.jp
}
}

\begin{document}
\maketitle

\begin{abstract}
High-performance Byzantine Fault Tolerant (BFT) blockchains are designed to achieve high throughput and low latency, yet their observed block time distributions often reveal complex behaviors arising from networking, pipelining, and deployment heterogeneity. In this paper, we diagnose HotStuff-based high-performance BFT consensus by modeling block times through a quorum-based multicast framework that links each block interval to quorum formation latency. We capture multimodal block time distributions using mixture models, where each component represents a distinct network condition characterized by effective transfer rate of block information. The proposed model is fitted to the bulk of mainnet block time data, while tail decay is analyzed separately to assess asymptotic behavior. Applying this methodology to Hyperliquid and Aptos mainnets, we find that Hyperliquid is well explained by a unimodal distribution, consistent with a relatively homogeneous validator deployment. In contrast, Aptos exhibits persistent multimodal structure and a pronounced shift following a consensus upgrade, reflecting heterogeneous deployments and diverse communication paths. These results demonstrate that mixture modeling of block time provides a practical and informative diagnostic tool for analyzing and monitoring high-performance BFT consensus.

\end{abstract}


\section{Introduction}
The evolution of blockchain consensus can be viewed as a continuous exploration of trade-offs among consistency, availability, and partition tolerance in distributed systems~\cite{brewer2000,gilbert2002}. Bitcoin's Nakamoto consensus introduced probabilistic finality through proof-of-work, where safety emerges asymptotically as blocks accumulate~\cite{nakamoto}. Ethereum's transition to proof-of-stake marked a paradigm shift toward deterministic finality, with its Gasper protocol combining Casper FFG and LMD-GHOST to achieve accountable safety under partial synchrony~\cite{gasper}. This trajectory has now advanced to high-performance Layer~1 blockchains, including Aptos, Sui, Hyperliquid, Solana, and Monad~\cite{aptos,sui,hyperliquid,solana,monadbft}. The research focus has shifted from proving safety alone to achieving instant finality with sub-second latency.

Underlying this performance evolution is the revival of Byzantine fault-tolerant (BFT) protocols in PoS contexts. Classical PBFT provides deterministic safety and liveness under partial synchrony, but incurs $O(n^2)$ communication overhead that limits scalability~\cite{pbft}. HotStuff addresses this bottleneck through threshold signatures and pipelined phases, enabling linear message complexity in the optimistic case~\cite{hotstuff}. Modern deployments further maximize throughput by decoupling payload dissemination from consensus ordering and adopting aggressive pipelining. Among certificate-driven HotStuff variants, Hyperliquid and Aptos represent prominent production systems: Hyperliquid optimizes specifically for a decentralized perpetual exchange with strict latency requirements~\cite{hyperliquid,hyperliquidNode}, while Aptos targets a broader ecosystem of general-purpose decentralized applications~\cite{aptos}.

Among the observable metrics in deployed BFT systems, block time stands out as a practical diagnostic signal. Defined as the interval between consecutive block proposals, it is directly obtainable from on-chain data and reflects how information disseminates among validators. In practice, BFT consensus advances once a quorum forms, and this quorum-based completion corresponds to an order statistic.

High-performance BFT systems often exhibit multimodal block time distributions whose structure differs qualitatively from timeout-induced multimodality observed in Tendermint. In Tendermint protocols, timeouts create discrete peaks at integer multiples of the timeout interval. In contrast, HotStuff-based pipelines operating below the timeout threshold produce densely clustered peaks within a narrow range. Validator heterogeneity has a particularly sensitive impact on quorum completion time. This motivates mixture models that decompose the aggregate distribution into components representing distinct network conditions.

Existing analytical work on block time models the full-committee broadcast completion time, yielding Gumbel extreme-value asymptotics~\cite{fujihara2024,fujihara2025,fujihara2024brains}. In practice, certificate-driven BFT consensus advances once a quorum of votes forms, not when all nodes are informed. The present work shifts the analytical target from the broadcast maximum to the quorum completion time, derives its exact closed-form distribution, extends it to finite mixtures that capture validator heterogeneity, and validates the resulting framework against production mainnet data.

This paper presents a diagnostic framework for high-performance BFT consensus, with empirical validation on mainnet data from Hyperliquid and Aptos. We model quorum formation time using a quorum multicast completion-time process and perform fitting on the bulk of the distribution together with a separate tail analysis. By fitting block time distributions with mixture models, we quantitatively identify the fraction of validators contributing to consensus delay under each network condition, providing actionable insights for system tuning. The main contributions of this study are:

\begin{itemize}
 \item We derive an exact closed-form distribution for quorum completion time in HotStuff-style pipelined BFT systems, shown in Eq.~(\ref{eq:pdf_M_threshold}) using a continuous-time Markov chain, and extend it to finite mixture models that capture deployment heterogeneity. 
 The model provides a principled link between quorum formation latency and observable block time distributions. 
 \item We validate the proposed model using production mainnet block time data from Hyperliquid and Aptos. By fitting unimodal and mixture regimes and analyzing tail behavior, we demonstrate how real-world block time distributions reflect mixed operating conditions and structural shifts induced by protocol updates.
 \item We find that fitted mixture parameters and tail decay serve as effective diagnostic signals for high-performance BFT deployments. Exponential tails and unimodal fits indicate homogeneous validator operation under a single effective rate, while multimodality and heavy tails diagnose heterogeneity in validator performance. These diagnostics provide quantitative guidance for validator management, infrastructure optimization, and monitoring consensus behavior over time.
\end{itemize}

The paper is organized as follows. Section~\ref{sec:related} reviews related work on BFT systems and latency modeling. Section~\ref{sec:method} details the datasets and measurement procedures, formalizes block time as a quorum completion time, and outlines our fitting procedure. Section~\ref{sec:theory} presents the theoretical analysis, including the closed-form distribution and moment formulas, with extensions to parallel and serial compositions. Section~\ref{sec:results} reports empirical fits on Hyperliquid and Aptos. Section~\ref{sec:discussion} discusses interpretation, implications, and limitations. Section~\ref{sec:conclusion} concludes.

\begin{figure*}[t]
\centering
\resizebox{0.95\textwidth}{!}{%
\begin{tikzpicture}[
    font=\sffamily,
    thick,
    >=Latex,
    timeline/.style={very thick, color=blue!60!black},
    divider_thick/.style={dashed, ultra thick, color=blue!60!black},
    divider_thin/.style={dashed, thin, color=blue!40!black!60},
    msg_arrow_red/.style={->, color=red, thin},
    msg_arrow_black/.style={->, color=black, thin},
    msg_arrow_blue/.style={->, color=blue, thin},
    black_arrow/.style={->, color=black, thin},
    phase_label/.style={align=center, font=\bfseries},
    block_node/.style={
        draw=black!85, 
        very thick, 
        rounded corners=8pt, 
        minimum height=1.1cm, 
        minimum width=2.0cm, 
        font=\sffamily\large\bfseries, 
        align=center,
        inner sep=5pt
    }
]

    \def\yClient{4}
    \def\yLeader{3}
    \def\yFOne{2}
    \def\yFTwo{1}
    \def\yFThree{0}

    \def\xStart{0}
    \def\xReqArrival{2.5}   
    \def\xPreStart{3.5}     
    \def\xPreEnd{6.5}       
    \def\xVoteEnd{9.5}      
    \def\xCommitStart{10.5} 
    \def\xCommitEnd{13.5}   
    \def\xNewVoteEnd{16.5}  
    \def\xEnd{17.0}         

    \foreach \y/\name in {
        \yClient/Client,
        \yLeader/Leader\\(Validator0),
        \yFOne/Follower1\\(Validator1),
        \yFTwo/Follower2\\(Validator2),
        \yFThree/Follower3\\(Validator3)
    } {
        \draw[timeline] (\xStart, \y) -- (\xEnd, \y);
        \node[align=left, font=\small, anchor=north west] at (0, \y) {\name};
    }

    \draw[divider_thin] (\xReqArrival, \yFThree - 0.9) -- (\xReqArrival, \yClient + 1);
    \foreach \x in {\xPreStart, \xPreEnd, \xVoteEnd, \xCommitStart, \xCommitEnd, \xNewVoteEnd} {
        \draw[divider_thick] (\x, \yFThree - 0.9) -- (\x, \yClient + 1);
    }

    \node[phase_label, black, anchor=south] at ({(\xStart+\xReqArrival)/2 }, \yClient) {REQUEST\\(PROPOSAL)};
    \node[phase_label, red, anchor=south] at ({(\xPreStart+\xPreEnd)/2}, \yClient+0.2) {PREPARE};
    \node[phase_label, red, anchor=south] at ({(\xPreEnd+\xVoteEnd)/2}, \yClient+0.2) {VOTE};
    \node[phase_label, red, anchor=south] at ({(\xCommitStart+\xCommitEnd)/2}, \yClient) {COMMIT\\\textcolor{blue}{PREPARE}};
    \node[phase_label, blue, anchor=south] at ({(\xCommitEnd+\xNewVoteEnd)/2}, \yClient+0.2) {VOTE};

    \draw[black_arrow] (1.5, \yClient) -- (\xReqArrival, \yLeader);
    
    \foreach \ySource in {\yLeader, \yFOne, \yFTwo, \yFThree} {
        \foreach \yTarget in {\yLeader, \yFOne, \yFTwo, \yFThree} {
            \ifx\ySource\yTarget \else
                \ifnum\ySource=\yLeader
                    \draw[msg_arrow_red] (\xPreStart, \ySource) -- (\xPreEnd-0.1, \yTarget);
                \else
                    \draw[msg_arrow_black] (\xPreStart, \ySource) -- (\xPreEnd-0.1, \yTarget);
                \fi
            \fi
        }
    }
    \foreach \ySource in {\yLeader, \yFOne, \yFTwo, \yFThree} {
        \foreach \yTarget in {\yLeader, \yFOne, \yFTwo, \yFThree} {
            \ifx\ySource\yTarget \else
                \draw[msg_arrow_black] (\xPreEnd+0.1, \ySource) -- (\xVoteEnd-0.1, \yTarget);
            \fi
        }
    }
    \foreach \ySource in {\yLeader, \yFOne, \yFTwo, \yFThree} {
        \foreach \yTarget in {\yLeader, \yFOne, \yFTwo, \yFThree} {
            \ifx\ySource\yTarget \else
                \ifnum\ySource=\yLeader
                    \draw[msg_arrow_blue] (\xCommitStart+0.1, \ySource) -- (\xCommitEnd-0.1, \yTarget);
                \else
                    \draw[msg_arrow_black] (\xCommitStart+0.1, \ySource) -- (\xCommitEnd-0.1, \yTarget);
                \fi
            \fi
        }
    }
    \foreach \ySource in {\yLeader, \yFOne, \yFTwo, \yFThree} {
        \foreach \yTarget in {\yLeader, \yFOne, \yFTwo, \yFThree} {
            \ifx\ySource\yTarget \else
                \draw[msg_arrow_black] (\xCommitEnd+0.1, \ySource) -- (\xNewVoteEnd-0.1, \yTarget);
            \fi
        }
    }

    \draw[<->, red, thick] (\xReqArrival + 0.05, \yFThree - 0.8) -- (\xPreStart - 0.05, \yFThree - 0.8) 
        node[midway, below, font=\scriptsize] {$T_{\mathrm{create}}$};
    \draw[<->, red, thick] (\xPreStart + 0.05, \yFThree - 0.8) -- (\xPreEnd - 0.05, \yFThree - 0.8) 
        node[midway, below, font=\scriptsize] {$T_{\mathrm{broadcast}}$};
    \draw[<->, red, thick] (\xPreEnd + 0.05, \yFThree - 0.8) -- (\xVoteEnd - 0.05, \yFThree - 0.8) 
        node[midway, below, font=\scriptsize] {$T_{\mathrm{validate}}$};
    \draw[<->, blue, thick] (\xVoteEnd + 0.05, \yFThree - 0.8) -- (\xCommitStart - 0.05, \yFThree - 0.8) 
        node[midway, below, font=\scriptsize] {$T_{\mathrm{create}}$};
    \draw[<->, blue, thick] (\xCommitStart + 0.05, \yFThree - 0.8) -- (\xCommitEnd - 0.05, \yFThree - 0.8) 
        node[midway, below, font=\scriptsize] {$T_{\mathrm{broadcast}}$};
    \draw[<->, blue, thick] (\xCommitEnd + 0.05, \yFThree - 0.8) -- (\xNewVoteEnd - 0.05, \yFThree - 0.8) 
        node[midway, below, font=\scriptsize] {$T_{\mathrm{validate}}$};

    \def\yBlock{-3.0} 

    \node[block_node, fill=red!20, anchor=west] (b1) at (\xReqArrival, \yBlock) {Block $r$};

    \node[block_node, fill=blue!20, anchor=west] (b2) at (\xVoteEnd, \yBlock) {Block $r{+}1$};

    \draw[->, very thick, >=Latex] (b2.west) -- (b1.east);

    \draw[->, very thick, >=Latex] (\xNewVoteEnd, \yBlock) -- (b2.east);

    
    \draw[->, thick, >=Latex] (\xReqArrival + 0.05, -1.2) -- (\xReqArrival + 0.05, \yBlock + 0.55)
        node[midway, right, font=\small] {$ts_{h}$};

    \draw[->, thick, >=Latex] (\xVoteEnd + 0.05, -1.2) -- (\xVoteEnd + 0.05, \yBlock + 0.55)
        node[midway, right, font=\small] {$ts_{h+1}$};

\end{tikzpicture}
}
\caption{Pipeline communication phases in high-performance BFT.}
\label{fig:three_phases}
\end{figure*}
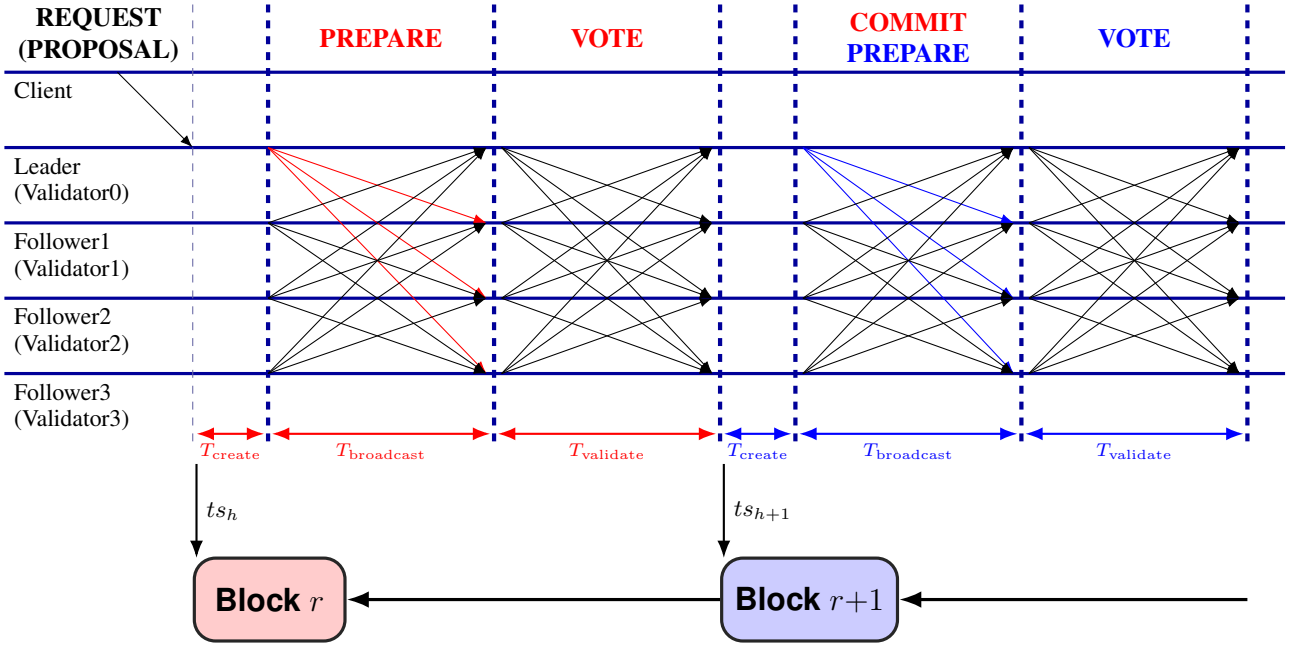

\section{Related Work}
\label{sec:related}
Classical PBFT established the three-phase baseline for Byzantine fault-tolerant agreement with a canonical three-message-delay latency under the standard Byzantine resilience assumption~\cite{pbft,kuznetsov2021fast}. Subsequent systems explored scalability, communication overhead, and performance trade-offs in blockchain contexts, including Tendermint, which is widely used in public PoS networks~\cite{tendermint}.

HotStuff addresses the quadratic communication bottleneck by having the leader aggregate votes into quorum certificates, which results in linear communication in the steady state and enables pipelining~\cite{hotstuff}. This pipeline design increases throughput by overlapping phases across heights, but under the 3-chain commit rule, block commit latency spans seven message delays~\cite{libra}. Facebook's Libra (later Diem) adopted this as LibraBFT~\cite{libra}, and in 2021, 2-chain variants such as Jolteon (deployed as DiemBFT v4) reduced commit latency from 7 to 5 message delays by committing in two rounds rather than three, at the cost of quadratic communication during view changes~\cite{jolteon,diembft}. To mitigate quadratic view changes due to failed leaders, DiemBFT v4 introduces a leader reputation mechanism that biases leader selection away from underperforming nodes.

The Diem project was terminated in 2022, but its consensus ideas and the Move execution environment were inherited by Aptos~\cite{aptos}. In mid-2023, Aptos deployed Quorum Store to decouple payload dissemination from ordering via a Narwhal-inspired mempool mechanism~\cite{narwhal,aptosQS}, drawing on the DAG-based BFT paradigm where Bullshark demonstrated low-latency consensus without additional communication overhead~\cite{bullshark}. Jolteon$^\ast$ (Order Votes, AIP-89) further reduces common-case consensus latency from 5 to 3 message delays by replacing leader-aggregated vote collection with an all-to-all vote exchange~\cite{aip89}. Order Votes optimizes the consensus path, whereas Quorum Store and Baby Raptr (BR) optimize the payload path through batching and decoupled dissemination~\cite{raptr,aptos_babyraptr_blog}. BR's optimistic data availability mechanism illustrates how protocol design can create multiple latency paths within a single system. We provide a message delay interpretation aligned with Fig.~\ref{fig:three_phases} in Section~\ref{sec:method}.

Several recent studies analyze BFT consensus latency through analytical modeling. Chan et al.\ derive expected consensus time for IBFT and HotStuff under datacenter topologies~\cite{chan2025consensus}. Hwerbi et al.\ derive round duration distributions for Narwhal under Gaussian propagation assumptions~\cite{hwerbi2025narwhal}. Gai et al.\ model quorum vote collection as an order statistic under Gaussian RTT in their Bamboo framework~\cite{gai2021dissecting}. These works focus on expected latency or assume homogeneous conditions, and validate primarily against simulation rather than production data.

Prior work analyzes validator broadcast latency under a full-committee abstraction and uses extreme value theory to obtain a Gumbel limit, motivating block time models built as convolutions of Gumbel distributions~\cite{fujihara2024}.
Appendix~B of that work briefly remarks that progress in BFT deployments is gated by forming a Byzantine quorum, but does not derive a systematic model for quorum completion time.
Subsequent work develops formulas for single and parallel broadcast time~\cite{fujihara2025,fujihara2024brains}.
This line of work remains centered on broadcasting to all nodes, whereas practical certificate-driven consensus advances once the quorum forms.
A related theoretical direction formalizes the size-synchrony antagonism in PoS finality time, showing that larger committees face inherently greater difficulty in 
maintaining synchronous communication~\cite{fujihara2025icbc}.

\section{Method}
\label{sec:method}
\subsection{Block time decomposition}
Consider a committee of $N$ validator nodes connected by a peer-to-peer network, where up to $f$ nodes can be Byzantine. Under standard $3f{+}1$ resilience, progress requires $M=2f{+}1$ votes.
For a given round $r$, a designated leader proposes a candidate block $B_r$ and the committee exchanges authenticated messages to order and commit the block.
As illustrated in Figure~\ref{fig:three_phases}, we abstract the leader period into proposal broadcast (\textsc{prepare}), quorum vote formation (\textsc{vote}), and commit/advance, with these phases merged over consecutive rounds. 

For blocks indexed by height $h$ with on-chain timestamps $ts_h$, the observed inter-block interval is $t_h = ts_{h+1} - ts_h$.
Following the decomposition in prior work~\cite{fujihara2025,fujihara2024brains}, block time in general blockchain systems consists of three parts:
\begin{equation}
T_{\mathrm{block}} = T_{\mathrm{create}} + T_{\mathrm{broadcast}} + T_{\mathrm{validate}},
\label{eq:Tblock_general}
\end{equation}
where $T_{\mathrm{create}}$ is the block creation time, $T_{\mathrm{broadcast}}$ is the proposal propagation time (\textsc{prepare} phase), and $T_{\mathrm{validate}}$ is the vote collection and QC formation time (\textsc{vote} phase).
Since $T_{\mathrm{create}} \ll T_{\mathrm{broadcast}} + T_{\mathrm{validate}}$ 
in pipelined BFT systems, the block time random variable $T_h$ satisfies
\begin{align}
T_h &\approx T_{\mathrm{broadcast}} + T_{\mathrm{validate}} \label{eq:block_decomp}\\
    &= T_M + T_{\parallel}, \label{eq:block_serial_decomp}
\end{align}
where $T_M$ is the quorum multicast completion time (the time for a proposal to reach $M$ validators) and $T_{\parallel}$ is the parallel completion time (the maximum over concurrent multicasts).

This block time $t_h$ is distinct from consensus latency. 
In Jolteon$^\ast$ pipelines~\cite{aip89}, once a quorum certificate (QC) 
for Block $h$ is formed at the end of the \textsc{vote} phase, replicas 
broadcast commit-certificate votes (CC-votes) and advance to the next round.
Thus, $t_h$ captures the cadence at which the pipeline advances to the 
next proposal, while finality latency includes the additional CC-vote 
collection phase.
The size-synchrony trade-off formalized in~\cite{fujihara2025icbc} provides additional context for interpreting block time differences between small and large validator sets.
Fig.~\ref{fig:three_phases} overlays two consecutive heights to illustrate 
pipelining: red arrows indicate the latency path for the current block, 
while blue arrows indicate phases of the next block that overlap the 
\textsc{commit} phase.

\subsection{Quorum multicast model}
\label{sec:ctmc}

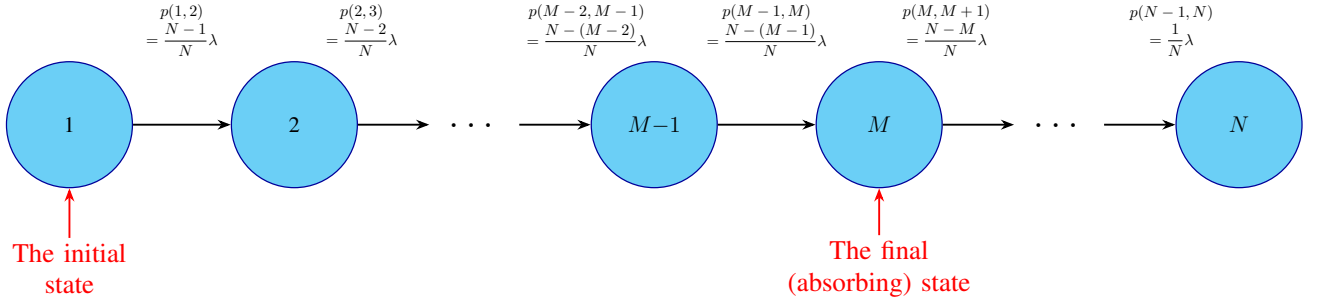
\begin{figure*}[t]
\centering
\resizebox{0.95\textwidth}{!}{%
\begin{tikzpicture}[
  >=Stealth,
  state/.style={
    circle,
    draw=blue!60!black,
    fill=cyan!50,
    line width=0.8pt,
    minimum size=28mm,
    inner sep=0pt
  }
]

\node[state] (s1)  at (0,0)   {\Large 1};
\node[state] (s2)  at (5,0)   {\Large 2};

\coordinate (dotsLeft) at (8,0);
\coordinate (dotsRight) at (10,0);

\node[state] (sMm1) at (13,0) {\Large $M{-}1$};
\node[state] (sM)   at (18,0) {\Large $M$};

\coordinate (dots2Left) at (21,0);
\coordinate (dots2Right) at (23,0);

\node[state] (sN)   at (26,0) {\Large $N$};

\draw[->, very thick] (s1) -- (s2);
\draw[->, very thick] (s2) -- (dotsLeft);
\draw[->, very thick] (dotsRight) -- (sMm1);
\draw[->, very thick] (sMm1) -- (sM);

\draw[->, very thick] (sM) -- (dots2Left);
\draw[->, very thick] (dots2Right) -- (sN);

\node[font=\Huge] at ($(dotsLeft)!0.5!(dotsRight)$) {$\dots$};
\node[font=\Huge] at ($(dots2Left)!0.5!(dots2Right)$) {$\dots$};


\node[align=center] at ($(s1)!0.5!(s2)+(0,2.1)$)
  {$p(1,2)$\\[2pt]$=\dfrac{N-1}{N}\lambda$};

\node[align=center] at ($(s2)!0.5!(dotsLeft)+(0,2.1)$)
  {$p(2,3)$\\[2pt]$=\dfrac{N-2}{N}\lambda$};

\node[align=center] at ($(dotsRight)!0.5!(sMm1)+(0,2.1)$)
  {$p(M-2,M-1)$\\[2pt]$=\dfrac{N-(M-2)}{N}\lambda$};

\node[align=center] at ($(sMm1)!0.5!(sM)+(0,2.1)$)
  {$p(M-1,M)$\\[2pt]$=\dfrac{N-(M-1)}{N}\lambda$};

\node[align=center] at ($(sM)!0.5!(dots2Left)+(0,2.1)$)
  {$p(M,M+1)$\\[2pt]$=\dfrac{N-M}{N}\lambda$};

\node[align=center] at ($(dots2Right)!0.5!(sN)+(0,2.1)$)
  {$p(N-1,N)$\\[2pt]$=\dfrac{1}{N}\lambda$};

\node[align=center, text=red, font=\LARGE] (init) at ($(s1)+(0,-3.2)$)
  {The initial\\state};
\draw[->, red, line width=1.2pt] (init.north) -- (s1.south);

\node[align=center, text=red, font=\LARGE] (final) at ($(sM)+(0,-3.2)$)
  {The final\\(absorbing) state};
\draw[->, red, line width=1.2pt] (final.north) -- (sM.south);

\end{tikzpicture}%
}
\caption{State transition view of the quorum multicast completion time.}
\label{fig:state_transition}
\end{figure*}

We model quorum formation as a completion-time process: the leader broadcasts to all $N-1$ peers, but progress requires only $M-1$ responses. This $M$-of-$N$ completion criterion is referred to as quorum multicast.
To formalize $T_M$ in the above decomposition, consider a continuous-time Markov chain (CTMC) with states 
$i=1,2,\ldots,M,\ldots,N$ (Figure~\ref{fig:state_transition}), where state $i$ 
represents that exactly $i$ validators have received the proposal.
The process starts at $i=1$ (the leader) and transitions to $i{+}1$ 
(up to the absorbing threshold $M$) when one new validator becomes informed.
Each uninformed validator independently receives the proposal at a 
normalized rate $\lambda/N$. When $i$ validators already hold the 
proposal, the aggregate transition rate to state $i{+}1$ is
\begin{equation}
\alpha_i = \frac{N-i}{N}\lambda, \qquad i=1,2,\ldots,M-1,
\label{eq:alpha_i}
\end{equation}
where $\lambda>0$ is an effective transfer rate parameter.
This formulation is the continuous-time limit of the coupon-collector 
broadcast model with geometric sojourn times~\cite{fujihara2025,fujihara2024brains}.
For standard quorum $M=\lceil 2N/3\rceil$, the chain visits states $1,2,\ldots,M{-}1$ sequentially, giving $T_M$ as the sum of independent exponential sojourn times, each $X_i$ following $\mathrm{Exp}(\alpha_i)$:
\begin{equation}
T_M = \sum_{i=1}^{M-1} X_i.
\label{eq:TM_sum}
\end{equation}
Section~\ref{sec:theory} derives the closed-form distribution of $T_M$ 
and establishes its connection to block time through serial-parallel 
composition and mixture extensions.

\subsection{Datasets and measurement procedure}
\label{sec:data}

We use Hyperliquid data from the public AWS S3 export and Aptos data from the public GraphQL endpoint. Table~\ref{tab:datasets} summarizes the measurement windows. $N$ is set to the number of distinct proposers observed in each window. Block times are computed as consecutive timestamp differences. To exclude timeout-triggered view changes, the analysis range is capped at $t \le 0.4$\,s for Hyperliquid (inferred from histogram separation) and $t < 1.0$\,s for Aptos (aligned with the configured \texttt{round\_initial\_timeout\_ms}). The pre-BR Aptos window is shorter because approximately one million samples provide sufficient statistical power for distribution fitting and comparative analysis.

\subsection{Parameter fitting with likelihood and mixtures}
For a processed sample set $\mathcal{D}=\{t_1,\dots,t_n\}$ with known committee size $N$, we treat $t_i$ as approximately independent draws from the $T_M$ distribution in Eq.~(\ref{eq:pdf_M_threshold}), which functions as a proxy for the overall latency (see theoretical justification in Section~\ref{sec:serial_composition}).
The fitting strategy adapts to the data structure:
\begin{itemize}
    \item{Unimodal:} For distributions with a single dominant mode, we estimate a single effective rate $\lambda$ via Maximum Likelihood Estimation (MLE).
    \item{Multimodal:} For distributions with multiple peaks, we use an Expectation-Maximization (EM) algorithm to estimate mixture weights $w_k$ and component rates $\lambda_k$.
\end{itemize}
In unimodal fitting, we estimate a single rate $\lambda$ by maximizing the log-likelihood:
\begin{equation}
\ell(\lambda)=\sum_{i=1}^{n}\log f_{T_M}(t_i\mid N,M_0,\lambda),
\label{eq:loglik}
\end{equation}
and obtain $\hat{\lambda}$ by a one-dimensional bounded optimization over $\lambda>0$.
$\lambda$ is initialized by matching the sample mean $\overline{t}$ to the 
theoretical mean $\E[T_M]$:
\begin{equation}
\lambda_{\mathrm{mom}}(M_0)
=
\frac{N}{\overline{t}}
\sum_{k=N-M_0+1}^{N-1}\frac{1}{k}.
\label{eq:lambda_mom}
\end{equation}
In multimodal fitting, we fit a finite mixture of $K$ components under the same fixed quorum $M_0$, where $K$ is the number of local maxima detected in the histogram:
\begin{equation}
p(t) = \sum_{k=1}^{K} w_k\, f_{T_M}(t\mid N,M_0,\lambda_k).
\end{equation}
$\{(w_k,\lambda_k)\}_{k=1}^K$ is estimated by EM.
In the E-step, responsibilities $r_{ik}$ are computed by normalizing $w_k\, f_{T_M}(t_i\mid N,M_0,\lambda_k)$ over $k$.
In the M-step, weights update as $w_k \leftarrow \frac{1}{n}\sum_i r_{ik}$, and each $\lambda_k$ is updated by weighted maximum likelihood:
\begin{equation}
\lambda_k \leftarrow \arg\max_{\lambda>0} \sum_i r_{ik}\,\log f_{T_M}(t_i\mid N,M_0,\lambda).
\end{equation}
The mixture is initialized by partitioning sorted samples into $K$ quantile groups, then fitting each group with the unimodal routine. Body fitting covers the full analysis range to capture dominant operating conditions. Tail fitting is performed separately using samples above the 90th percentile to characterize decay shape.

\begin{table}[t]
\caption{Dataset summary and measurement windows.}
\label{tab:datasets}
\centering
{\footnotesize
\setlength{\tabcolsep}{3pt}
\renewcommand{\arraystretch}{0.95}
\begin{tabularx}{\columnwidth}{l l >{\raggedright\arraybackslash}X r}
\toprule
Chain & Window (UTC) & Heights & Samples \\
\midrule
Hyperliquid & 2025-08-17--2025-08-26 & 700M--710M & 10.0M \\
Aptos (pre-BR) & 2025-03-07--2025-03-09 & 300M--301M & 1.2M \\
Aptos (post-BR) & 2025-08-06--2025-08-19 & 400M--410M & 10.0M \\
\bottomrule
\end{tabularx}
}
\vspace{-0.8em}
\end{table}

\section{Theoretical Analysis}
\label{sec:theory}

Prior broadcast models characterize the time for a message to reach all $N$ nodes as a maximum of propagation delays, leading to extreme value distributions such as Gumbel~\cite{fujihara2024}. In contrast, quorum-based progress requires only $M$ responses and corresponds to the $(M{-}1)$-th order statistic rather than the maximum. This shift from extreme-value asymptotics to order statistics yields a hypoexponential distribution (a sum of exponentials with distinct rates) whose tail decays as a single exponential in Eq.~(\ref{eq:tail_asymp}) rather than the doubly exponential form characteristic of Gumbel.

\subsection{Closed-form distribution and tail asymptotics}

Since $T_M$ is the sum of $M{-}1$ independent exponentials with distinct rates~(Eq.~\ref{eq:TM_sum}), its probability density involves a convolution of exponential distributions. Taking the Laplace transform converts this convolution into a product:
\begin{equation}
F(s)=\prod_{i=1}^{M-1}\frac{1}{\dfrac{Ns}{(N-i)\lambda}+1},
\qquad (2 \le M \le N).
\label{eq:laplace_F}
\end{equation}
To obtain the time-domain distribution from~\eqref{eq:laplace_F}, we apply partial-fraction expansion. Each pole in $F(s)$ corresponds to an exponential component in the density. The pole locations $-\alpha_i = -\frac{(N-i)\lambda}{N}$ for $i=1,\ldots,M-1$ are the departure rates defined in Eq.~\eqref{eq:alpha_i}. The smallest rate $\alpha_{M-1}=\frac{(N-M+1)\lambda}{N}$ governs the far-tail decay.
Rewriting each factor in shifted-pole form and separating the constant prefactor:
\begin{align}
F(s)
&=\prod_{i=1}^{M-1}\frac{1}{\frac{Ns}{(N-i)\lambda}+1}
=\prod_{i=1}^{M-1}\frac{(N-i)\lambda/N}{s+\alpha_i}
\notag\\
&=\underbrace{\frac{\lambda^{M-1}(N-1)!}{N^{M-1}(N-M)!}}_{C}
\prod_{i=1}^{M-1}\frac{1}{s+\alpha_i}.
\label{eq:Fs_factored}
\end{align}
The constant $C$ captures the time-scale dependence on $\lambda$, while the product of shifted poles determines the distributional shape.
Applying partial-fraction decomposition to the pole product:
\begin{align}
\prod_{i=1}^{M-1}\frac{1}{s+\alpha_i}
&=\sum_{i=1}^{M-1}\frac{A_i}{s+\alpha_i},
\notag\\
\text{where}\quad
A_i
&=\frac{1}{\prod_{j\neq i}(\alpha_j-\alpha_i)}
\notag\\
&=\left(\frac{N}{\lambda}\right)^{M-2}
\frac{(-1)^{M-1-i}}{(i-1)!(M-1-i)!}.
\label{eq:PFD}
\end{align}
The inverse transform of each term is $A_i e^{-\alpha_i t}$. Combining this with the prefactor $C$ and substituting the closed-form expressions results in the probability density function:
\begin{align}
f_{T_M}(t)
&=C\sum_{i=1}^{M-1}A_i\,e^{-\alpha_i t}
\notag\\
&=\frac{\lambda}{N}\frac{(N-1)!}{(N-M)!}
\sum_{i=1}^{M-1}\frac{(-1)^{M-1-i}}{(i-1)!(M-1-i)!}
\notag\\
&\quad\times
\exp\!\left[-\frac{(N-i)\lambda}{N}t\right].
\label{eq:pdf_expansion}
\end{align}
This expression is a weighted sum of $M-1$ exponential terms. Each term reflects contributions from paths through the chain that experience different combinations of sojourn times before reaching the quorum threshold. The alternating signs arise from the partial-fraction residues.
When the $N-1$ receiver delays are independent $\mathrm{Exp}(\lambda/N)$ variables, $T_M$ is the $(M-1)$-th order statistic with density:
\begin{equation}
\begin{aligned}
f_{T_M}(t)
&=\binom{N-1}{M-1}\frac{M-1}{N}\lambda
\exp\!\left(-\frac{(N-M+1)\lambda}{N}t\right)
\\
&\qquad\times
\left(1-\exp\!\left(-\frac{\lambda}{N}t\right)\right)^{M-2}.
\end{aligned}
\label{eq:pdf_M_threshold}
\end{equation}

The cumulative distribution function admits a closed form through a change of variables. Define $u = 1-\exp(-\lambda t/N)$, representing the fraction of nodes informed by time $t$. Under independent exponential delays, $u$ follows a Beta distribution with parameters $(M-1, N-M+1)$, giving the regularized incomplete beta function form:
\begin{equation}
F_{T_M}(t)=I_u(M-1,N-M+1).
\label{eq:cdf_beta}
\end{equation}
Because $f_{T_M}(t)$ is a finite mixture of exponentials, the slowest-decaying component dominates at large $t$. As $t\to\infty$, the term with the smallest exponent $\alpha_{M-1}$ gives
\begin{equation}
f_{T_M}(t) \simeq e^{-\Lambda t},
\quad
S_{T_M}(t) \simeq C_{\mathrm{tail}}\,e^{-\Lambda t}
\qquad (t\to\infty),
\label{eq:tail_asymp}
\end{equation}
where $\Lambda=\alpha_{M-1}$ and $C_{\mathrm{tail}}>0$ depends on the mixture coefficients. This single-exponential tail contrasts with the doubly exponential form characteristic of Gumbel and provides the theoretical baseline for the empirical analysis in Section~\ref{sec:results}.

\subsection{Moments of quorum completion time}
\label{sec:moments}

Since $T_M$ is the sum of $M-1$ independent exponential sojourn times with rates $\alpha_i$, its first two moments follow directly from the additivity of means and variances.
The mean and variance are obtained by summing $1/\alpha_i$ and $1/\alpha_i^2$ over $i=1,\ldots,M-1$, respectively:
\begin{align}
\E[T_M]
&=\sum_{i=1}^{M-1}\frac{1}{\alpha_i}
=\sum_{i=1}^{M-1}\frac{N}{(N-i)\lambda}
=\frac{N}{\lambda}\sum_{j=1}^{M-1}\frac{1}{N-j}
\notag\\
&=\frac{N}{\lambda}\bigl(H_{N-1}-H_{N-M}\bigr),
\label{eq:ETM}
\end{align}
\begin{align}
\Var(T_M)
&=\sum_{i=1}^{M-1}\frac{1}{\alpha_i^2}
=\frac{N^2}{\lambda^2}\sum_{j=1}^{M-1}\frac{1}{(N-j)^2}
\notag\\
&=\frac{N^2}{\lambda^2}\bigl(H^{(2)}_{N-1}-H^{(2)}_{N-M}\bigr),
\label{eq:VarTM}
\end{align}
with $H_n=\sum_{r=1}^{n}r^{-1}$ the $n$-th harmonic number and $H^{(2)}_n=\sum_{r=1}^{n}r^{-2}$ its order-two generalization.
The harmonic structure in both expressions arises from summing reciprocal powers of the decreasing departure rates as the chain approaches the quorum threshold.

For large committees with fixed quorum fraction $\rho=M/N$, the harmonic sums admit logarithmic approximations. Using $H_n\approx\ln n+\gamma$ for large $n$:
\begin{equation}
\E[T_M] \approx \frac{N}{\lambda}\ln\frac{1}{1-\rho}, 
\qquad
\mathrm{SD}(T_M) \approx \frac{\sqrt{N}}{\lambda}\sqrt{\frac{\rho}{1-\rho}}.
\label{eq:moments_asymp}
\end{equation}
The mean scales linearly with $N$, while the standard deviation scales as $\sqrt{N}$, so the coefficient of variation decreases for larger committees.

\subsection{Extension to parallel broadcast}
\label{sec:parallel_extension}

In the \textsc{vote} phase, all $N$ validators broadcast simultaneously, and the phase completes when the slowest finishes. This parallel structure means the phase duration is governed by the maximum completion time:
\begin{equation}
T_{\parallel}=\max_{1\le j\le N} T_M^{(j)},
\label{eq:Tparallel_def}
\end{equation}
where $T_M^{(j)}$ is the quorum completion time for the $j$-th validator. We treat these as independent and identically distributed, which is justified when individual broadcast durations are short relative to network congestion timescales. The probability that all broadcasts complete by time $t$ equals the product of individual completion probabilities:
\begin{align}
\Pr(T_{\parallel}\le t)
&=\prod_{j=1}^{N}\Pr(T_M^{(j)}\le t)
=F_{T_M}(t)^N,
\notag\\
\Pr(T_{\parallel}>t)
&=1-F_{T_M}(t)^N
=1-\bigl(1-S_{T_M}(t)\bigr)^N.
\label{eq:Tparallel_cdf}
\end{align}
To bound this tail probability, we apply the union bound: the event that at least one broadcast exceeds $t$ is contained in the union of individual exceedance events. This gives
\begin{equation}
\Pr(T_{\parallel}>t)\le \min\!\bigl\{1,\, N S_{T_M}(t)\bigr\}.
\label{eq:union_bound}
\end{equation}
Since $S_{T_M}(t)$ decays exponentially with rate $\Lambda=\alpha_{M-1}$ from Eq.~\eqref{eq:tail_asymp}, we have the uniform bound $S_{T_M}(t)\le C e^{-\Lambda t}$ for some constant $C>0$ depending on $N$ and $M$.
The expected parallel completion time can be estimated using the tail-integral formula for non-negative random variables:
\begin{equation}
\E[T_{\parallel}] = \int_{0}^{\infty}\Pr(T_{\parallel}>t)\,dt.
\label{eq:tail_integral}
\end{equation}
Substituting the union bound and evaluating the integral by splitting at $t_0=\max\{0,\ln(NC)/\Lambda\}$:
\begin{align}
\E[T_{\parallel}]
&\le \int_{0}^{\infty}\min\!\bigl\{1,\, N C e^{-\Lambda t}\bigr\}\,dt
\notag\\
&= t_0 + \int_{t_0}^{\infty}NCe^{-\Lambda t}\,dt
= t_0 + \frac{1}{\Lambda}
\notag\\
&\le \frac{1}{\Lambda}\Bigl(\ln(NC)+1\Bigr).
\label{eq:ETparallel_bound}
\end{align}
Comparing with the mean from Eq.~\eqref{eq:ETM}, the parallel phase adds only $O(\log N)$ to the expected completion time:
\begin{equation}
\E[T_{\parallel}]=\E[T_M]+O(\log N).
\label{eq:parallel_overhead}
\end{equation}
For fixed quorum fraction $\rho=M/N$, the dominant rate $\Lambda=(1-\rho)\lambda$ remains $\Theta(1)$ while $\E[T_M]=\Theta(N)$, so the logarithmic overhead is asymptotically negligible.

\subsection{Phase composition}
\label{sec:serial_composition}

The block time $T_h$ is the sum of $T_M$ and $T_{\parallel}$. Since the two phases occur sequentially, 
\begin{equation}
\E[T_h] \approx \E[T_M]+\E[T_{\parallel}].
\label{eq:block_mean}
\end{equation}
For variance, additivity requires independence between phases. In practice, network conditions may induce positive correlation (e.g., congestion slowing both phases), so treating the phases as independent yields a conservative lower bound:
\begin{equation}
\Var(T_h) \approx \Var(T_M)+\Var(T_{\parallel}).
\label{eq:block_var}
\end{equation}
The moments of $T_M$ are given by Eqs.~\eqref{eq:ETM}--\eqref{eq:VarTM}. For $T_{\parallel}$, Eq.~\eqref{eq:ETparallel_bound} shows that the parallel phase contributes an additional $O(\log N)$ term beyond $\E[T_M]$. The variance of $T_{\parallel}$ can be computed from the second moment:
\begin{equation}
\E[T_{\parallel}^2] = 2\int_{0}^{\infty} t\,\bigl(1-F_{T_M}(t)^N\bigr)\,dt.
\label{eq:Tparallel_second_moment}
\end{equation}
Combining these results, the total block time has mean $\E[T_h]=\Theta(N)$ and standard deviation $\mathrm{SD}(T_h)=\Theta(\sqrt{N})$ for fixed quorum fraction $\rho$. The coefficient of variation $\mathrm{SD}(T_h)/\E[T_h]=\Theta(N^{-1/2})$ decreases with committee size, indicating that larger committees produce more predictable block times.
The exponential tail structure is preserved under serial composition: the sum of exponentially-tailed variables retains exponential decay with the same dominant rate $\Lambda$. 

\section{Empirical Results}
\label{sec:results}

\begin{figure*}[t]
\centering
\begin{subfigure}[t]{0.32\textwidth}
\centering
\includegraphics[width=\linewidth]{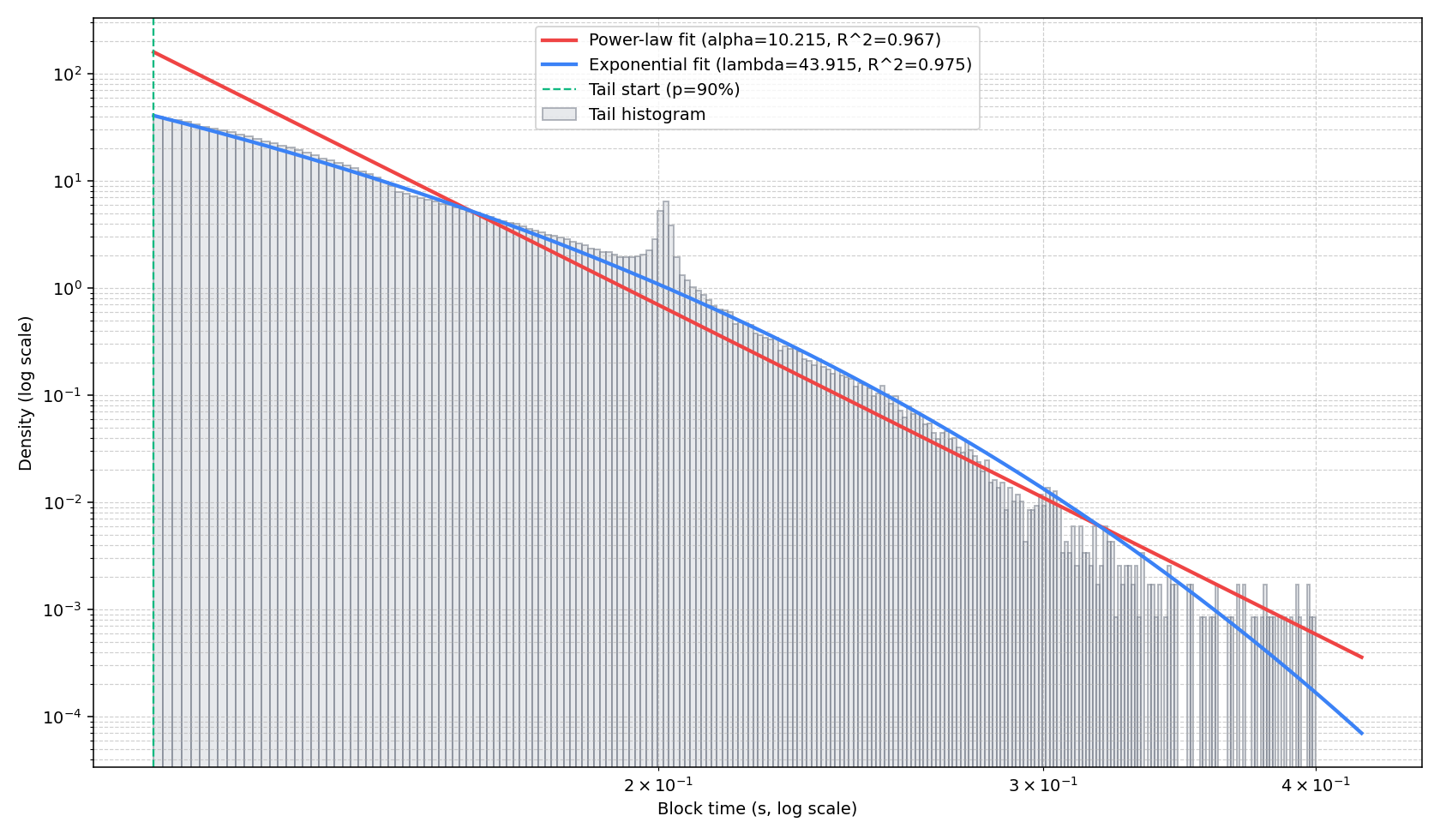}
\caption{Hyperliquid: tail shape analysis above the cutoff.}
\label{fig:hyperliquid_tail}
\end{subfigure}
\hfill
\begin{subfigure}[t]{0.32\textwidth}
\centering
\includegraphics[width=\linewidth]{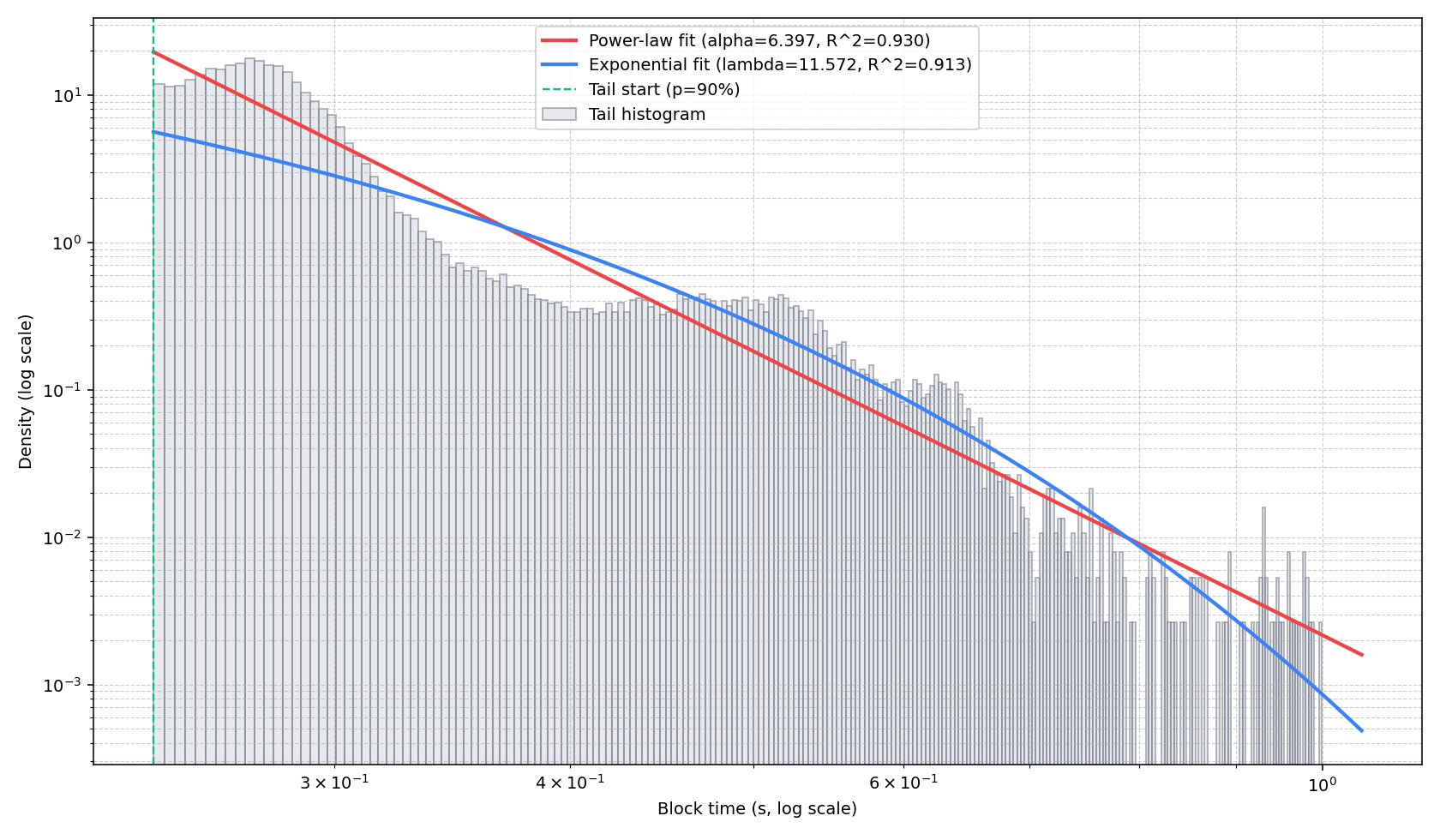}
\caption{Aptos pre-BR: tail shape analysis above the cutoff.}
\label{fig:aptos_tail_pre}
\end{subfigure}
\hfill
\begin{subfigure}[t]{0.32\textwidth}
\centering
\includegraphics[width=\linewidth]{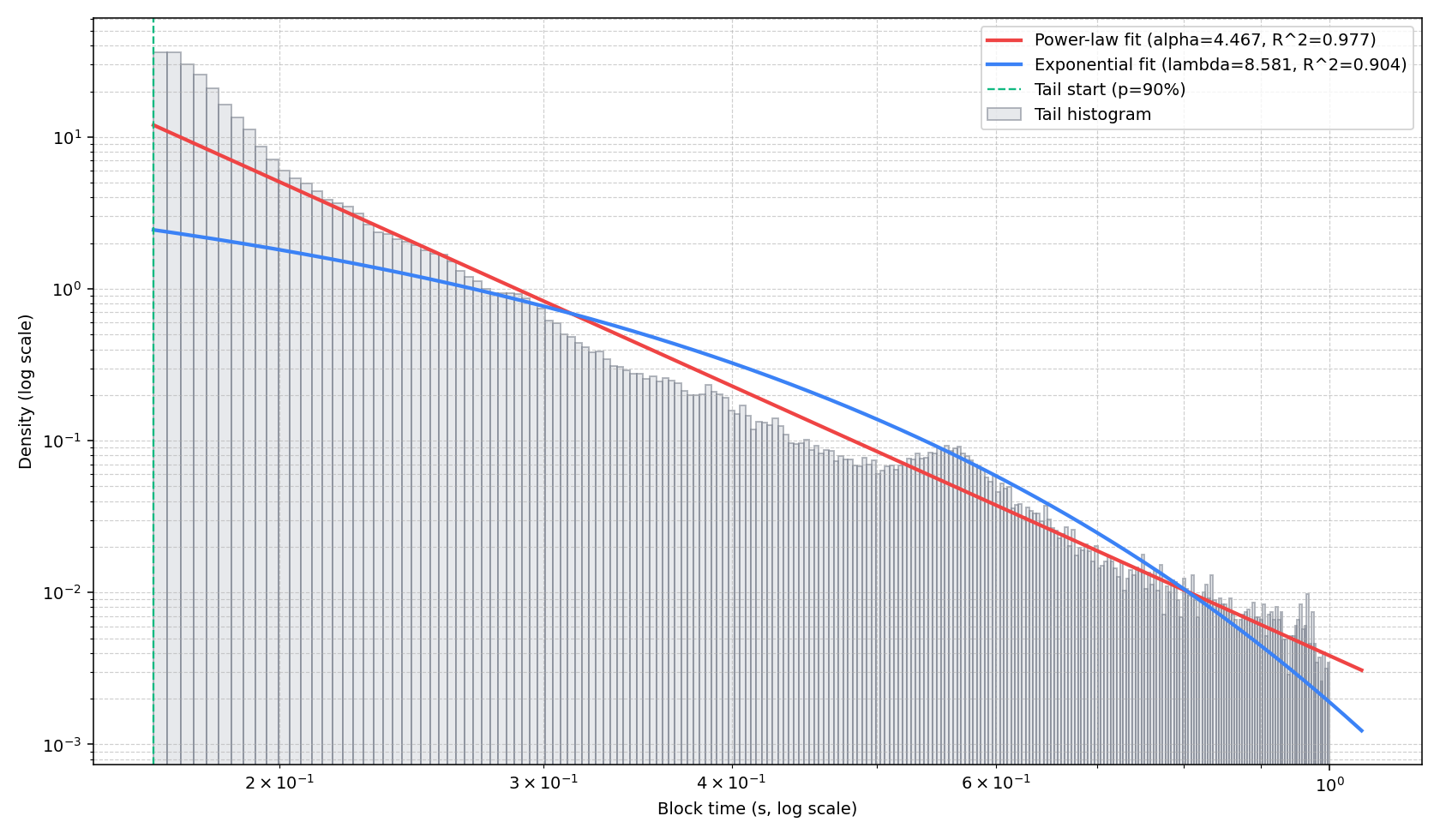}
\caption{Aptos post-BR: tail shape analysis above the cutoff.}
\label{fig:aptos_tail_post}
\end{subfigure}
\caption{Tail shape analysis of block time distributions for Hyperliquid and Aptos (pre-BR/post-BR).}
\label{fig:tail_diagnostics}
\end{figure*}

\subsection{Baseline validation on Hyperliquid}
Hyperliquid exhibits a unimodal block time distribution.
In Fig.~\ref{fig:mle_fits}(a), the histogram shows a single dominant peak near 0.06--0.07\,s with a narrow spread and smoothly decaying right tail.
The distribution is well captured by a unimodal fit ($K{=}1$), with estimated rate $\hat{\lambda}=267.6\,\mathrm{s}^{-1}$ (Table~\ref{tab:params}).

Tail decay analysis (Fig.~\ref{fig:tail_diagnostics}) compares exponential and power-law fits above the 90th percentile.
The exponential model achieves $R^2{=}0.975$ with tail decay rate $\hat{\Lambda}_{\mathrm{tail}}{=}43.9\,\mathrm{s}^{-1}$, compared to $R^2{=}0.967$ for power-law.
This preference for exponential decay is consistent with the theoretical CTMC baseline (Eq.~\eqref{eq:tail_asymp}), where a single effective rate $\lambda$ produces exponential tail decay.
The agreement supports the interpretation of homogeneous network conditions across the validator set.

\subsection{Mixture analysis on Aptos}
\begin{figure*}[t]
\centering
\begin{subfigure}[t]{0.32\textwidth}
\centering
\includegraphics[width=\linewidth]{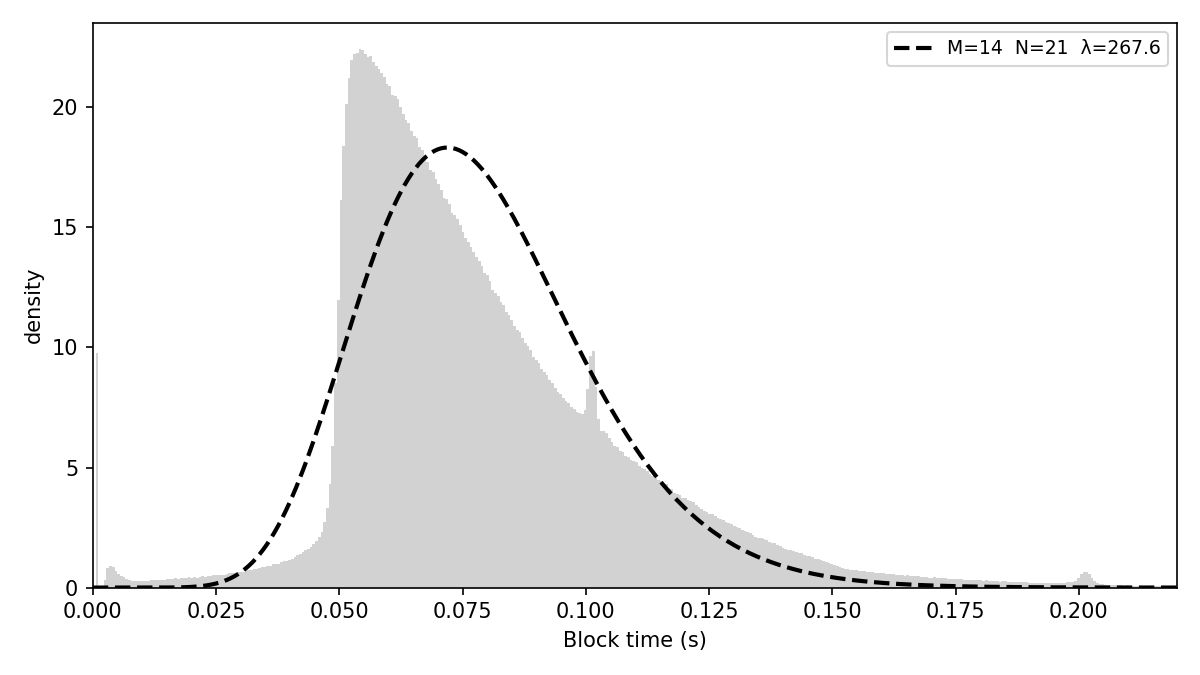}
\caption{Hyperliquid (Aug.): body fit (unimodal, $K{=}1$), fixed quorum $M_0{=}14$, $N{=}21$.}
\label{fig:hyperliquid_window_fit}
\end{subfigure}
\hfill
\begin{subfigure}[t]{0.32\textwidth}
\centering
\includegraphics[width=\linewidth]{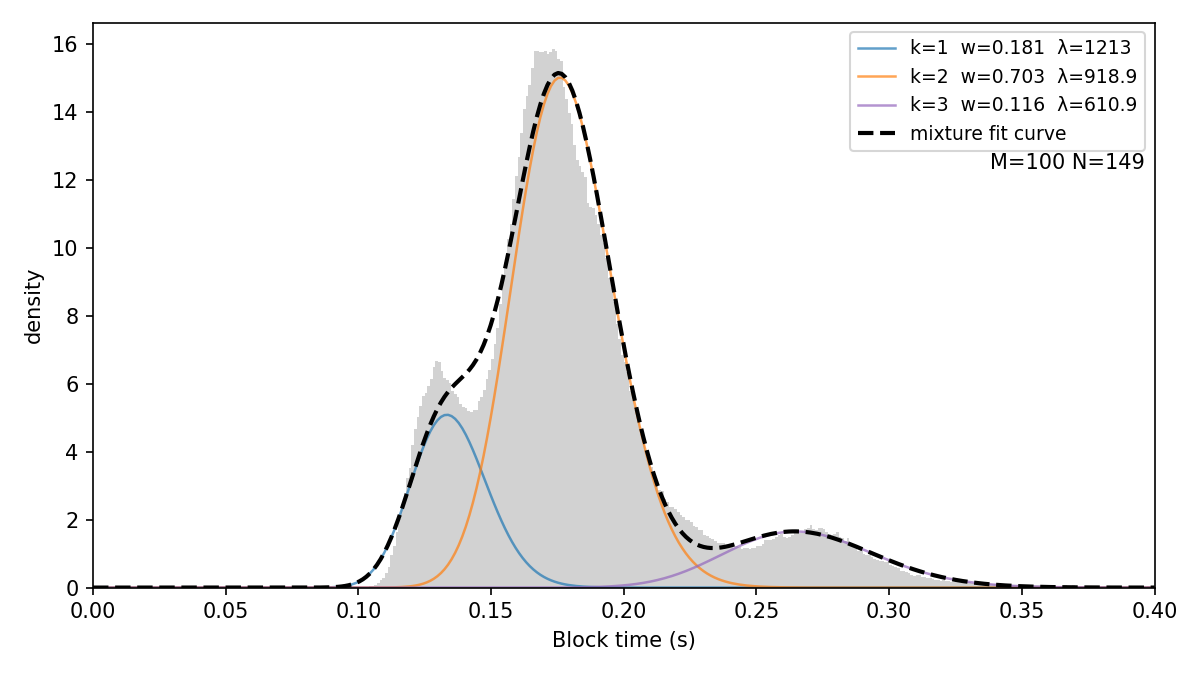}
\caption{Aptos (Mar., pre-BR): body fit ($K{=}3$ mixture), fixed quorum $M_0{=}100$, $N{=}149$.}
\label{fig:aptos_pre_br_peak_fits}
\end{subfigure}
\hfill
\begin{subfigure}[t]{0.32\textwidth}
\centering
\includegraphics[width=\linewidth]{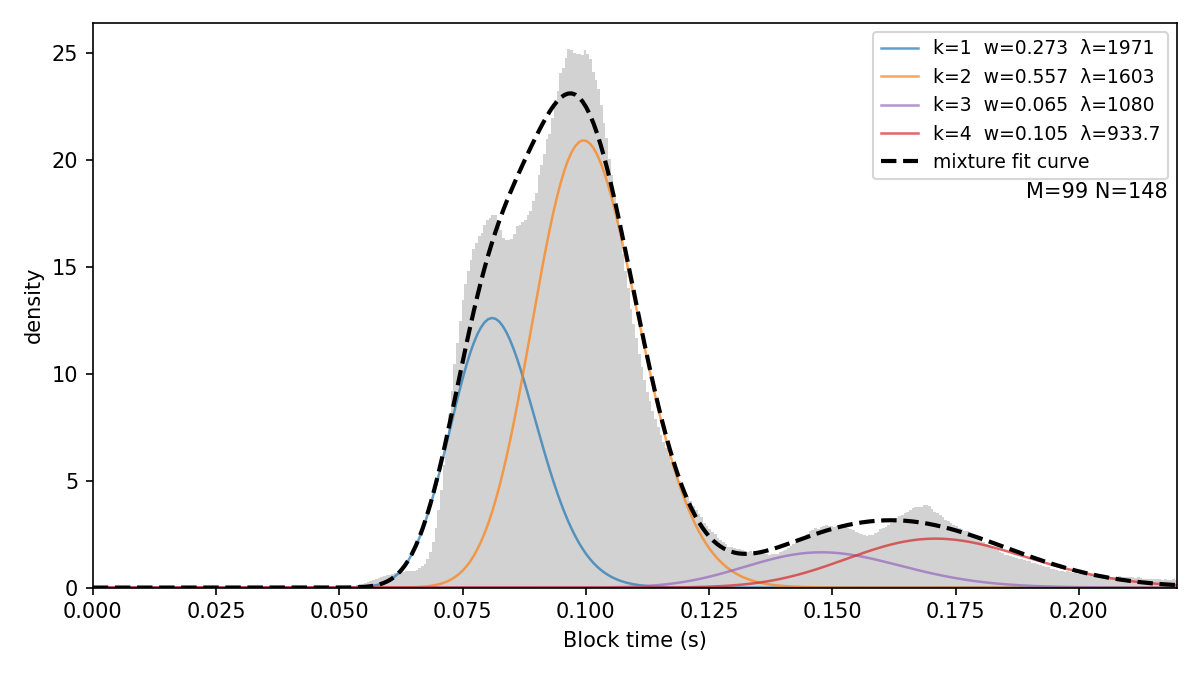}
\caption{Aptos (Aug., post-BR): body fit ($K{=}4$ mixture), fixed quorum $M_0{=}99$, $N{=}148$.}
\label{fig:aptos_post_br_peak_fits}
\end{subfigure}
\caption{Global exact model fits for block time under the fixed quorum constraint $M_0=\lceil 2N/3\rceil$.}
\label{fig:mle_fits}
\end{figure*}

In contrast, Aptos exhibits clear multimodality in Fig.~\ref{fig:mle_fits}(b)--(c), requiring mixture models to capture the heterogeneity of a larger validator set.
In the pre-BR period, the distribution (Fig.~\ref{fig:mle_fits}(b)) shows three distinct components.
Under the fixed quorum constraint ($M_0{=}100$), the fitted components partition validators into performance groups: a fast minority ($w_1{=}0.181$, $\lambda_1{=}1213\,\mathrm{s}^{-1}$), a majority group ($w_2{=}0.703$, $\lambda_2{=}919\,\mathrm{s}^{-1}$), and a slow component ($w_3{=}0.116$, $\lambda_3{=}611\,\mathrm{s}^{-1}$).
After the BR upgrade, two observable changes emerge (Fig.~\ref{fig:mle_fits}(c), Table~\ref{tab:params}).
First, all fitted rates increase: the fastest component rises from $1213$ to $1971\,\mathrm{s}^{-1}$ (+63\%), reflecting BR's latency reduction.
Second, the mixture structure shifts from three to four components, with the former slow component splitting into two sub-components ($\lambda_3{=}1080$, $\lambda_4{=}934\,\mathrm{s}^{-1}$).
The majority of rounds (approximately 56\%) now operate under the second component at $1603\,\mathrm{s}^{-1}$, while the two slower components together account for about 17\% of rounds.
The number of components $K$ is determined by histogram peak detection rather than information criteria: at these sample sizes (1.2M--10M blocks), BIC tends to favor additional components that lack physical interpretation, whereas peak-based selection identifies modes that correspond to distinct operating conditions.

Comparing tail decay across periods, the diagnostics (Fig.~\ref{fig:tail_diagnostics}) reveal departure from exponential behavior.
For pre-BR data, power-law fit achieves $R^2{=}0.930$ versus $R^2{=}0.913$ for exponential.
Post-BR shows a stronger power-law signature: $R^2{=}0.977$ versus $R^2{=}0.904$, with the gap widening from 0.017 to 0.073.
The power-law exponent also decreases ($\hat{\alpha}$: $6.40 \to 4.47$), indicating heavier tails after the upgrade.
This slower tail decay suggests the system does not operate under a single effective rate.

\section{Discussion}
\label{sec:discussion}

\subsection{Mixture parameters as diagnostics}
The mixture decomposition translates observed distributions into operational parameters: each $\lambda_k$ characterizes a network condition, and $w_k$ quantifies the fraction of rounds under that condition.
This granularity is not available from aggregate statistics.
Fitting before and after the BR upgrade tracks how each component's rate and weight evolved.

The parameters also suggest optimization levers.
Rate improvement targets $\lambda_k$ through infrastructure upgrades or protocol optimizations that reduce blocking delays.
Weight reduction targets $w_k$ by excluding slow validators via jailing or reducing their leader opportunities through reputation mechanisms.
The two act differently: rate improvement benefits all rounds in a given regime, while weight reduction shifts probability mass across regimes.

To quantify the impact of each lever, the aggregate expected block time can be derived from the mixture model.
From Eq.~\eqref{eq:ETM}, a single component with rate $\lambda$ contributes expected time $\frac{N}{\lambda}(H_{N-1}-H_{N-M})$ to the aggregate.
Since $N$ and $M$ are fixed for a given deployment, the system-specific harmonic factor $N(H_{N-1}-H_{N-M})$ acts as a constant multiplier.
Summing over all mixture components gives the aggregate expected block time:
\begin{equation}
\mathbb{E}[T_h] = N(H_{N-1}-H_{N-M}) \sum_{k} \frac{w_k}{\lambda_k}.
\label{eq:aggregate_mean}
\end{equation}
This expression shows that each component's contribution to latency is proportional to $w_k/\lambda_k$: components with low rates contribute disproportionately even when their weights are small.

In Aptos post-BR, the slow components ($k{=}3,4$) represent only 17\% of rounds by weight but account for 26\% of the aggregate latency, revealing a clear optimization target.
Upgrading or excluding these components yields approximately 10\% latency reduction; the similar magnitude of both interventions reflects a fundamental constraint---the slow components' impact is bounded by their combined weight.

More broadly, these diagnostics address why the inherent performance of HotStuff cannot be fully realized, rather than proposing modifications to the protocol itself. The contrast between Hyperliquid ($N{=}21$, unimodal) and Aptos ($N{\approx}148$, multimodal) illustrates that as the validator set grows, maintaining homogeneous performance becomes fundamentally more difficult. Approaching HotStuff's ideal latency requires mechanisms that enforce validator homogeneity---latency-based jailing or reputation-weighted leader selection---which directly target the $w_k$ and $\lambda_k$ that the mixture framework quantifies.

\subsection{Cross-system design comparison}

In validator management, Hyperliquid enforces strict latency discipline via a jailing mechanism~\cite{hyperliquidStaking}.
Validators may vote to jail peers that do not respond with adequate latency.
Jailed validators are excluded from consensus and receive no rewards until they self-unjail, creating strong economic incentives for low-latency infrastructure.
Aptos, in contrast, uses a reputation mechanism that reduces leader opportunities for underperforming validators~\cite{diembft,aptos} but does not exclude them from the voting quorum.
This design preserves fault tolerance at the cost of latency uniformity.
From a modeling perspective, jailing primarily reduces $w_{\mathrm{slow}}$ by excluding persistently slow validators, while reputation adjusts the distribution of $w_k$ without removal.

At the protocol level, Aptos decouples payload dissemination from consensus ordering through Quorum Store, a Narwhal-inspired DAG mechanism~\cite{aptosQS}.
This architecture creates path-dependent latency.
Validators with cached batches can vote immediately, but those missing data must fetch before voting.
BR's optimistic data availability amplifies this divergence~\cite{raptr}: leaders propose metadata without availability proofs, and validators experiencing cache misses incur additional message delays on the critical path.
This blocking fetch mechanism contributes to the heavier tails observed in Fig.~\ref{fig:tail_diagnostics}.
Additionally, Aptos conducts protocol upgrades via gradual rollout, so version heterogeneity among validators may have contributed to mode separation.
Hyperliquid's unimodal distribution and exponential tail suggest a more streamlined pipeline with less path divergence.
In the mixture framework, these architectural differences appear as additional components or increased $w_{\mathrm{slow}}$, rather than uniform degradation of all $\lambda_k$.

Infrastructure differences compound this effect. Hyperliquid recommends Tokyo-region placement and 200\,ms RTT targets~\cite{hyperliquidNode}; the majority of mainnet seed nodes are located in this region, and the small committee size ($N{=}21$) further limits path diversity. Aptos spans multiple regions and cloud providers with a much larger validator set ($N{\approx}148$), introducing substantially greater RTT variation~\cite{raptr} that directly affects the $\lambda_k$ of corresponding validator groups (Table~\ref{tab:params}).

\begin{table}[t]
\caption{Mixture fit parameters ($M_0=\lceil 2N/3\rceil$).$^{\dagger}$}
\label{tab:params}
\centering
{\footnotesize
\setlength{\tabcolsep}{4pt}
\renewcommand{\arraystretch}{1.0}
\begin{tabular}{l c c r r}
\toprule
Dataset & $(N, M_0)$ & $k$ & $w_k$ & $\lambda_k$ ($\mathrm{s}^{-1}$) \\
\midrule
Hyperliquid (Aug.) & (21, 14) & 1 & 1.000 & 267.6 \\
\midrule
\multirow{3}{*}{Aptos (pre-BR)} & \multirow{3}{*}{(149, 100)} 
  & 1 & 0.181 & 1213 \\
& & 2 & 0.703 & 918.9 \\
& & 3 & 0.116 & 610.9 \\
\midrule
\multirow{4}{*}{Aptos (post-BR)} & \multirow{4}{*}{(148, 99)} 
  & 1 & 0.273 & 1971 \\
& & 2 & 0.557 & 1603 \\
& & 3 & 0.065 & 1080 \\
& & 4 & 0.105 & 933.7 \\
\bottomrule
\multicolumn{5}{l}{\scriptsize $^{\dagger}$Components ordered by decreasing $\lambda_k$.}
\end{tabular}
}
\vspace{-0.8em}
\end{table}

\subsection{Limitations}

This analysis abstracts complex protocol logic into a quorum multicast primitive and relies on timestamp-derived block times.
Internal blocking events such as payload fetch delays or signature aggregation bottlenecks are not directly observable from on-chain data.
The fitted parameters should be interpreted as diagnostic indicators, and separating mixing effects from other sources of delay would require protocol-specific instrumentation.
The number of mixture components is determined empirically by peak detection, which may be sensitive to gradual parameter drift or transient network conditions.
The tail classification currently relies on $R^2$ comparison between exponential and power-law fits; formal goodness-of-fit testing remains a direction for future work.

The CTMC derivation assumes independent exponential delays across validators and across rounds.
In practice, cloud-region clustering, persistent validator slowness, and shared network bottlenecks introduce spatial and temporal correlation.
The independent model serves as a diagnostic baseline. Under independence, Eq.~\eqref{eq:tail_asymp} predicts exponential tail decay. The power-law signatures observed in Aptos deviate from this prediction, signaling the presence of correlated or path-dependent mechanisms.
Positive correlation reduces the effective number of independent delays, causing the quorum order statistic to shift right and broaden relative to Eq.~\eqref{eq:pdf_M_threshold}; equivalently, Eq.~\eqref{eq:cdf_beta} overestimates the completion probability at a given time.
The mixture extension partially absorbs spatial correlation at a coarse level, since validators experiencing correlated slowdowns appear collectively as a low-rate component.
Quantifying the residual bias requires comparison with simulation under explicit correlation structures, which is left to future work.

\section{Conclusion}
\label{sec:conclusion}
This paper develops a quorum-based model for block time distributions in high-performance BFT systems.
Block time is linked to quorum formation latency and is modeled as a quorum completion time with an exact closed-form distribution derived from a continuous-time Markov chain. Likelihood-based fitting provides a coarse alignment between theoretical predictions and observed distributions, with mixture decomposition identifying distinct network conditions and tail shape diagnosing rate homogeneity.
The approach provides a common interface for comparing implementations, detecting mixed operating paths, and monitoring protocol upgrades using timestamp-derived block times.
On our datasets, Hyperliquid exhibits a unimodal distribution with near-exponential tail decay, consistent with its homogeneous validator deployment. Aptos exhibits persistent multimodal structure and a clear shift after the BR rollout, reflecting validator heterogeneity and diverse communication paths.
Fitted mixture parameters help localize heterogeneity in effective transfer rates and payload availability paths, and provide quantitative reference points for guiding improvements.
Future work includes tracking additional chains or time windows for continuous monitoring, change-point detection, and joint bulk-tail modeling to reduce hyperparameter sensitivity while preserving interpretability.

\end{document}